___

BRIEF
COMMUMCATIONS

"Gas-Dynamic" Expansion of a Fast-Electron Flux in a Plasma

V. N. Mel'nik
Institute of Radioastronomy, Ukrainian Academy of Sciences, Khar'kov, Ukraine

During solar activity, spacecraft record fast electrons [1] that move from the Sun along the magnetic field lines. The so-called solar radio bursts of type III are attributed to these electrons [1, 2]. An adequate model of propagation of fast electron fluxes through a plasma must be constructed to explain the obtained experimental data [3]. In constructing such a model, it is important to incorporate both the interaction of fast electrons with plasma waves, which are excited precisely by these electrons, and the propagation effects. Note that, for a laboratory plasma, such problems have already been considered using an alternative formulation. Thus, Fainberg and Shapiro [4] studied the spatial structure of the plasma turbulence that arises during the injection of a fast electron beam into a semibounded plasma; and Bespalov and Trakhtengerts [5] investigated the characteristics of the isotropic expansion of a high-energy electron cloud. For our purposes, it is of most interest to consider the one-dimensional motion of the perturbation front due to the electron beam and its dependence on the boundary conditions. Such a formulation of the problem was first applied by Ryutov and Sagdeev [6], taking the quasi-linear equations of weak turbulence as a basis and using the approximation in which the time of quasi-linear relaxation is small compared to that of electron expansion. They formulated the "quasi-gas-dynamic" equations and found self-similar solutions for them.

In [7, 8], we derived a more complete system of "gas-dynamic" equations and found a self-similar solution for these equations in the problem of instant electron injection. Using these results, we constructed a model for the bursts of type III [9]. This model explains most of the properties of the bursts at large distances (which correspond to relatively low frequencies) from the source. At the same time, it is evident that, near the source, the time dependence of the source must be taken into account. Here, we solve the problem of one-dimensional motion of fast electrons through a plasma, taking into account the time dependence of an electron source and assuming the source to be monoenergetic.

Assume that, at the point x = 0, there is an electron source described by the distribution function

$$f_{bound}(v,t) = f(v)\psi(t) = f_0\delta(v - u_0)\psi(t), \quad t \geq 0 \tag{1}$$

In the case where $n_b / n \ll 1$ [where $n_b = \int f(v)dv$ and $n$ is the plasma density], the electron expansion is described by the quasi-linear equations [10, 11]

$$\frac{\partial f}{\partial t} + v\frac{\partial f}{\partial x} = \frac{4\pi^2 e^2}{m^2}\frac{\partial}{\partial v}\frac{W}{v}\frac{\partial f}{\partial v} \tag{2}$$

$$\frac{\partial W}{\partial t} = \frac{\pi\omega_p}{n}v^2 W\frac{\partial f}{\partial v}, \qquad \omega_p = kv \tag{3}$$

In equation (3), we neglected the group velocity of the plasmons, because $v \gg v_{Te}$ (where $v_{Te}$ is the thermal velocity of the electrons). The problem in which

$$\tau_e = \left(\omega_{pe} \frac{n_b}{n}\right)^{-1} \ll t \tag{4}$$

is of particular interest from the point of view of applications. Then, as shown in [7, 8], we can go from the kinetic description to the gas-dynamic one. In this case, the following electron distribution function $f_s(v,x,t)$ and spectral density of the plasma wave energy $W_s(v,x,t)$ are established at each point:

$$f_s = \begin{cases} p(x,t), & v < u(x,t) \\ 0, & v > u(x,t) \end{cases}$$

$$W_s = \begin{cases} W_0(x,t), & v < u(x,t) \\ 0, & v > u(x,t) \end{cases} \tag{5}$$

For the functions $p(x,t)$ and $u(x,t)$ we have the following gas-dynamic equations:

$$\frac{\partial}{\partial t} pu + \frac{1}{2}\frac{\partial}{\partial x} pu^2 = 0,$$

$$\frac{1}{2}\frac{\partial}{\partial t}(1+\beta)pu^2 + \frac{1}{3}\frac{\partial}{\partial x} pu^3 = 0,$$

$$\frac{1}{3}\frac{\partial}{\partial t}(1+\alpha)pu^3 + \frac{1}{4}\frac{\partial}{\partial x} pu^4 = 0 \tag{6}$$

where $\alpha(x,t)$ and $\beta(x,t)$ determine the level of Langmuir turbulence induced by fast electrons, that is, $\alpha = \int dk W_s / \int dv \frac{mv^2}{2} f_s$ and $\beta = \int dk \frac{k}{\omega_p} W_s / \int dv m v f_s$. Integration of (2) over the close neighborhood of $v = u$ (see, for example, [12]) yields the equation

$$\frac{\partial u}{\partial t} + u \frac{\partial u}{\partial x} = 0 , \tag{7}$$

which closes system (6).
To find $W_0$, we use the equation

$$\frac{\partial f_s}{\partial t} + v \frac{\partial f_s}{\partial x} = \frac{\omega_p}{m} \frac{\partial}{\partial v} \frac{1}{v^3} \frac{\partial W_s}{\partial t}, \tag{8}$$

This equation follows from equations (2) and (3) and, for v < u, takes the form

$$\frac{\partial p}{\partial t} + v \frac{\partial p}{\partial x} = \frac{\omega_p}{m} \frac{\partial}{\partial v} \frac{1}{v^3} \frac{\partial W_0}{\partial t} \tag{9}$$

Note, that Fainberg and Shapiro [4] used equation (8) to find the asymptotic behavior of the spectral energy density of plasma waves both close to and far from the plasma boundary.

When $v = u$, we find the boundary conditions for $W_0(v,x,t)$ in the following way. First, we integrate equation (8) over the velocities in the close neighborhood of $v = u$. Then, we perform the same integration for (8) multiplied by $v$. Thus, we find two conditions

$$\left.\frac{\partial W_0}{\partial t}\right|_{v=u} = 0, \tag{10}$$

$$\left.\frac{\partial u}{\partial t} W_0\right|_{v=u} = 0. \tag{11}$$

Hence, we have equations (6), (7), and (9) with the boundary conditions (10) and (11) for solving our problem in the gas-dynamic approximation.

Vedenov and Ryutov [12] showed that the arrival of the first particles at a certain point $x$ results in the establishment of $W_0$ with $W_0|_{v=u} \neq 0$. Thus, from condition (11), it follows that

$$\frac{\partial u}{\partial t} = 0 . \tag{12}$$

Accordingly, equation (7) gives

$$\frac{\partial u}{\partial x} = 0 \tag{13}$$

and, consequently,

$$u(x,t) = const . \tag{14}$$

Then, from equations (6) and (9) and conditions (10) and (11), we obtain

$$p(x,t) = p(t - 2x/u),$$

$$W_0(v,x,t) = \frac{m}{\omega_p} v^4 (1 - \frac{v}{u}) p(t - \frac{2x}{u}) + \frac{m}{\omega_p} v^3 \varphi(v,x) , \tag{15}$$

$$\beta(x,t) = 1/3 + 2p(-x/u)/3p(t - \frac{2x}{u}),$$

$$\alpha(x,t) = 1/2 + 3p(-x/u)/2p(t - \frac{2x}{u}).$$

Let us assume that the electrons move through a plasma with $W_T \ll W_0$ (where $W_T$ is the energy of thermal plasma fluctuations). Then, for the straight line x =ut, which corresponds to the fastest electrons, $W_0 \approx v^4$. This circumstance, along with (15), yields the following expression for g(x, v):

$$\varphi(x,v) = \frac{v^2}{u} p(-x/u). \tag{16}$$

To find an explicit form for p(y), we use the fact that, during the electron expansion in the half-space x > 0, the total number of fast electrons

$$\int_0^t dt \int_0^\infty dv f_{bound} = \int_0^{ut} dx \int_0^\infty dv f_s , \qquad (17)$$

the momentum

$$\int_0^t dt \int_0^\infty dv mv f_{bound} = \int_0^{ut} dx \left[ \int_0^\infty dv mv f_s + \int_0^\infty dk \frac{k}{\omega_p} W_s \right], \qquad (18)$$

and the energy

$$\int_0^t dt \int_0^\infty dv \frac{mv^2}{2} f_{bound} = \int_0^{ut} dx \left[ \int_0^\infty dv \frac{mv^2}{2} f_s + \int_0^\infty dk W_s \right] \qquad (19)$$

of the electron-plasmon system remain constant. From the system of integral equations (17) - (19), we find

$$p(y) = p(-y) = \frac{f_0}{u^2} \psi(y) \qquad (20)$$

The source function shown in Fig. 1 illustrates the obtained solution. Figure 2 shows the spatial and time variation of the height of the plateau. The maximal number of electrons are seen to move with a velocity $v = u/2$.

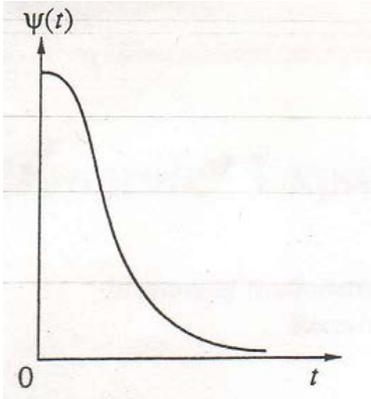

Fig. 1. Function of the electron source.

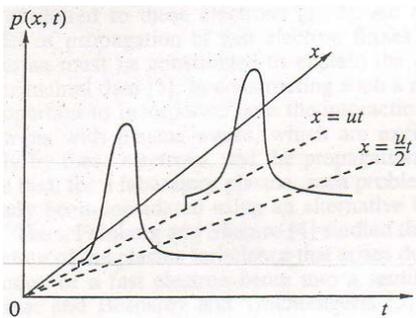

Fig. 2. Spatial and time variations of the height of the plateau.

According to expressions (15), the maximum spectral density of the plasmon energy corresponds to the straight line $x = ut/2$. Here, $\alpha \approx 1/2$ and $\beta \approx 1/3$, unlike the nonstationary case of quasi-linear electron-flux relaxation [12] in which $\alpha = 2$ and $\beta = 1$ (in our problem, these values correspond to the straight line $x = ut$). The concentration of fast electrons and plasmons on the straight line $x = ut/2$ is due to fact that the electrons with fairly high velocities with the distribution function $\partial f / \partial v > 0$ come ahead of the flux front ($x > ut/2$) at a given point $x$. In the fast processes of quasi-linear relaxation, a part of the energy is transferred from the electrons to the plasmons, which results in the electron slowing down. At the rear of the front ($x < ut/2$), the situation is the opposite: there are mostly electrons with relatively low velocities with the distribution functions $\partial f / \partial v < 0$. This results in plasmon energy absorption and, consequently, in electron acceleration. The accumulation of both electrons and plasmons occurs near $x = ut/2$. Fainberg and Shapiro [4] were first to draw attention to the effect of the "accumulation" of plasma waves in the problem of constant electron injection for a limited time. However, they did not study the structure of the accumulation region and its dependence on the plasma boundary conditions. The analysis carried out above shows, that the form of the electron source does not change with the electron expansion and the velocity of the main part of electrons is equal to half the maximum velocity.

The author would like to thank A.G. Boev and Ya.M. Sobolev for fruitful discussions.